# PVGIS approach for assessing the performances of the first PV grid-connected power plant in Morocco

**Abdelfettah BARHDADI\* and Mouncef BENNIS**

*Semiconductors Physics and Solar Energy Team (P.S.E.S.)*
*Ecole Normale Supérieure, University of Mohammed V-Agdal, Rabat, Morocco*
*P.O. Box: 5118, Takaddoum, Rabat 10000, Morocco*
*&*
*The Abdus Salam International Centre for Theoretical Physics, P.O. Box 586, Trieste, Italy*

**Abstract**

In this paper, we apply the PVGIS method for estimating the performance of the first grid-connected PV micro-power plant in Morocco. PVGIS approach provides analysis and assessment of in-site solar energy resources and predicts with good accuracy the potential of PV systems in term of electricity production. We find that annual total power generation of the micro-power is slightly higher than that initially expected at the installation stage and actually measured. The yearly predicted and measured power production values agree to about 2 %. However, individual monthly production can have larger discrepancy.

---

**\*** *Corresponding author (Senior Associate of the Abdus Salam ICTP)*
 *Phone: (212) 537 758 096 / 664 936 815*
 *Fax: (212) 537 759 063*
 *E-mails:* barhdadi@ictp.it  *or*  abdelbar@fsr.ac.ma

## I- Introduction:

With its high solar radiation and increasing need for energy, Morocco is set to become a future player in solar electricity production chain. Its strategic energy plan is in sync with the international trend which, in face of the growing energy demand and the global warming challenge, makes renewable energies and more particularly solar energy a priority. The main objective of the Moroccan integrated solar development plan is to set up until 2020, in five appropriate sites of the country, several power stations for electricity production from solar energy with a total capacity of 2000 MW. The two well developed technologies, Concentrated Solar Power (CSP) and Photovoltaic (PV), are designated to be used in these stations.

For PV technology, it is well known that system setting performances are strongly dependent on the climate conditions. The most important parameters influencing these performances are the solar radiation impinging at the surface of the PV modules and the ambient temperature that affect losses from these modules. The precise nature of this influence depends mainly on the PV technology used. A number of studies have been made to determine how the efficiency varies for different PV technologies [1, 2]. The result is a set of approximate expressions for module performance for given climatic characteristics. This is useful for a general appraisal of the technology. However, for a given PV installation obtaining the necessary climatic data is a non-trivial task, and so detailed performance evaluations become difficult. Yet the deviation of the conversion efficiency under real conditions from that obtained at Standard Test Conditions (STC) is large enough to have an influence on the economic feasibility of PV installations. Therefore it is important to be able to estimate the average performance of PV systems in sites where long-term detailed climatic data are not available. For this end, Suri et al [3-5] have developed at the Joint Research Centre of the European Commission their well established open source software named PhotoVoltaic Geographical Information System (PVGIS) which combines a long-term expertise from laboratory research, monitoring and testing with geographical knowledge. PVGIS is used as a research tool for the performance assessment of PV technology in geographical regions, and as a support system for policy-making in the European Union. A web interface was developed to provide interactive access to the data, maps and tools to other research and education institutes, decision-makers, PV professionals and system owners as well as to the general public.

The purpose of this paper is to apply PVGIS approach to the first Moroccan grid-connected micro-power PV plant built in Morocco with the aim to provide an analysis of in-site solar energy resources and PV electricity production from this station.

## II- Methodology:

PVGIS method uses the Geographical Information System (GIS) GRASS to combine geospatial data with known correlations for estimation of the performance of crystalline/thin film silicon modules under varying irradiance and temperature over large regions. The method has been successfully applied on the European and African continents, on the Mediterranean basin and on the South-West Asia region.

Applying this approach, the PV performance can be estimated for any geographical location in the area under investigation. Combined with a system for accessing the data interactively via the Internet, the results are made available for those interested.

The details of the PVGIS methodology and development can be found in key reference papers [6, 7]. We propose in the section to recall the essential.

**II-1- Performance of crystalline silicon PV systems:**

Most of terrestrial PV systems are made from mono-crystalline or poly-crystalline silicon materials. For these materials studies have been performed to calculate the system performances under varying conditions of irradiance and temperature. The starting point for performing this calculation is the set of semi-empirical formulas for the voltage V and current I of PV modules at maximum power point, given in [1] and used in modified form in [2]. From these we can calculate the relative conversion efficiency of the module ($\eta_{rel}$), i.e. the efficiency relative to the efficiency $\eta_0$ measured at STC:

$$\eta_{rel}(G_i, T_m) = \frac{I_m V_m}{I_{m,STC} V_{m,STC}} = \frac{G_i}{G_0}[1+\alpha_i(T_m-T_0)] \times \left[1 + c_1 \ln\frac{G_i}{G_0} + c_2\left(\ln\frac{G_i}{G_0}\right)^2 + \beta_v(T_m-T_0)\right] \quad (1)$$

Here $I_m$ and $V_m$ are respectively the real current and voltage values of the module at maximum power point at in-plane irradiance $G_i$ and module temperature $T_m$. $I_{m,STC}$ and $V_{m,STC}$ are the corresponding values, at STC of in-plane irradiance $G_0 = 1$ kW/m$^2$ and $T_0 = 25$ °C, at which the reference efficiency $\eta_0$ of PV modules is normally measured. The other parameters are empirical and were already determined experimentally in reference [2]. For mono-crystalline silicon modules which are the case of PV system considering in this work, the values of these parameters were found as: $\alpha_i = 1.20 \ 10^{-3}$ °C, $\beta_v = -4.60 \ 10^{-3}$ °C, $c_1 = 3.3 \ 10^{-2}$ and $c_2 = -9.2 \ 10^{-3}$. Different values would be needed for other PV technologies.

Since the module temperature $T_m$ is not known a priori, the following approximate relation between $T_m$, the ambient temperature $T_{amb}$ and the nominal operating cell temperature $T_{noct}$, provided by the module supplier, has been employed:

$$T_m = T_{amb} + (T_{noct} - 20)\frac{G_i}{800} \quad (2)$$

For modules in a free-standing configuration, an approximate value for $T_{noct}$ is normally given as $T_{noct} = 48$°C.

Given these relationships between PV module performance and meteorological conditions, we can calculate the power output of a given grid-connected crystalline silicon PV system with nominal peak power $P_{nom}$ as:

$$P(G_i, T_m) = P_{nom} \ \eta_{rel}(G_i, T_{amb})\frac{G_i}{G_0} \quad (3)$$

From equation 3, we see that in order to estimate the performance of a system at a given geographical location it is necessary to estimate both the irradiance in a given plane (in-plane irradiance) $G_i$ and the ambient temperature $T_{amb}$ at any given time during the day. This estimate must then be integrated over the day for a sufficient number of days during the year to obtain a good yearly average.

**II-2- Solar radiation database:**

The construction of high spatial resolution data sets for solar radiation has been previously reported [3, 4]. A brief description should therefore be sufficient here.

The computational approach is based on a solar radiation model *r.sun*, and the spline interpolation techniques *s.surf.rst* and *s.vol.rst* that are implemented within the open-source GIS software GRASS [8]. The *r.sun* model algorithm uses the equations published in the European Solar Radiation Atlas [9]. The model estimates direct, diffuse and reflected components of the clear-sky and real-sky global irradiance and/or irradiation for horizontal or inclined surfaces. The total daily irradiation (Wh/m$^2$) is computed by the integration of the irradiance values (W/m$^2$) that are calculated at a time step of 15 min from sunrise to sunset. For each time step, the computation accounts for sky obstruction (shadowing) by local terrain features, calculated from the digital elevation model.

The needed inputs for calculating average monthly and yearly irradiation were:

- Monthly averages of daily sums of global irradiation available at the meteorological ground stations [9];
- Monthly averages of the ratio of diffuse to global irradiation at the same ground stations (for irradiation on inclined planes);
- Monthly values of the Linke atmospheric turbidity [10];
- Digital Elevation Model (DEM) derived from SRTM-30 data [11].

The data from 566 meteorological ground stations are monthly averages of measurements over the period 1981-1990. Monthly averages of the clear-sky irradiation were calculated using the *r.sun*. Based on clear-sky irradiation and the ground station values of average real-sky solar irradiation, the clear-sky index was calculated for the ground station positions. Interpolating the clear-sky index and the ratio of diffuse to global irradiation makes it possible to develop spatial (GIS) databases of the average monthly real-sky irradiation on an arbitrarily inclined plane.

The grid resolution of DEM was chosen 1x1 km for two reasons. First, a higher-resolution DEM makes it possible to take account of the elevation in calculating the clear-sky irradiation. Secondly, when calculating the daily sum of irradiation, it becomes possible to take into account shadows from nearby terrain features. In mountainous regions this can be very significant.

Using *r.sun*, and taking into account shadowing effects from nearby terrain, we have obtained an estimate of the monthly and yearly averages of global irradiation/irradiance values at any inclination for any location. The relative RMS error of the developed database for monthly horizontal irradiation is within the interval of 3.2 to 7.8 % and the yearly average is 3.7 %. For more details of the calculation method, one may see Suri et al reference publications [3, 4].

**II-3- Temperature database:**

Temperature data come from approximately 800 ground stations over the period 1995-2003. These stations provide temperature measurements at 6:00, 9:00, 12:00, 15:00 and 18:00 GMT. The spatial distribution of the stations is heterogeneous with the highest density in Western Europe. For each station, the monthly and yearly averages of the temperature at the measurement time have been calculated. To estimate the temperature between measurement times, the known temperatures were fitted to a 2$^{nd}$ order polynomial. The coefficients of the polynomial were then interpolated separately over the geographical region using a multivariate regularized spline with tension *s.vol.rst*. The same grid resolution of the GIS data was used as for solar radiation (1x1 km). The RMS error from interpolation and fitting of monthly averages is in the range 0.5-0.7°C.

Thus, for any given point, coefficients of the polynomial to calculate the average temperature at any time during the day have been obtained. The methodology of developing the temperature database is described in detail in [5].

**II-4- Efficiency of silicon module:**

Given the average monthly real-sky irradiation and the polynomials for the daytime temperature a modified version of *r.sun* is used which incorporates the relative efficiency given by Equations (1) and (2) to integrate the daily PV energy output variation for crystalline silicon modules at any given location using Equation (3).

One thing to note when integrating Equation 3 using this method is that the irradiance $G_i$ used is the instantaneous irradiance for average meteorological conditions. For the linear term $G_i/G_0$ calculating $G_i$ using the average conditions is sufficiently accurate. However, for the calculation of $\eta_{rel}(G_i, T_{amb})$ it becomes necessary to take into account the actual values of irradiance. Most of the energy from a PV system is produced when the irradiance is close to clear-sky values. We have therefore chosen to use the clear-sky irradiance, $G_{i,cs}$, in this expression, whereby the equation for the power output becomes:

$$P(G_i, G_{i,cs}, T_m) = P_{nom}\, \eta_{rel}(G_{i,cs}, T_{amb})\frac{G_i}{G_0} \qquad (4)$$

The clear-sky and the real-sky irradiances, as well as the instantaneous value of the ambient temperature are calculated directly by the modified version of *r.sun* at 15 min. time step, and the expression for *P* is integrated over the day from sunrise to sunset.

**III- Solar power in Morocco:**

Solar power in Morocco is enabled by one of the highest rates of solar insulation of any country (~ 3000 hours per year of sunshine). Recently, Morocco has launched one of the world's largest solar energy projects costing an estimated $9 billion. The aim of the project is to generate 2000 MW of solar generation capacity by the year 2020. Five solar power stations are to be constructed. The Moroccan Agency for Solar Energy (MASEN), a public private venture, has been established to lead the project. MASEN has invited expressions of interest in the design, construction, operation, maintenance and financing of the first of the five planned solar power stations, the 500 MW plant in the southern town of Ouarzazate. This first plant will be commissioned in 2015 and the entire project in 2019. Once completed, the solar project will increase by 14% the contribution of solar energy in the annual electricity capacity and prevent the emission of 3.7 million tons of $CO_2$ yearly.

Supported by strong hydropower sources and the newly installed wind energy parks (147 MW installed and 975 MW under deployment). Morocco plans also a $13 billion expansion of wind, solar and hydroelectric power generation capacity and associated infrastructure that should see the country get 42% of its electricity from renewable sources by 2020. Moreover, Morocco is the only African country to have a power cable link to Europe. Therefore, it aims to benefit from the $573.8 billion expected to come from the ambitious pan-continental Desertec Industrial Initiative.

## IV- Identification of the first Moroccan PV micro-power plant:

On June 2007, the "Office National de l'Electricité" (ONE), which is the Moroccan Power Utility, launched a micro-solar PV station in Tit Mellil near Casablanca, to be the first of its kind in Africa. Named "Chourouk", which means Sunrise, this PV micro-power plant is part of an overall strategy set up by ONE to promote renewable energy forms and encourage people to use them. The technical characteristics of the plant are summarized in table 1.

| **First Moroccan PV micro-power plant** ||
|---|---|
| Location | Tit Mellil, Province of Mediouna, Grand Casablanca, Morocco |
| Latitude | 33° 33' 28" North |
| Longitude | 7° 29' 8" West |
| Elevation above sea level | 116 m |
| PV modules | 1024 PV panels of 45 Wp each installed on the roof of ONE building that houses the Regional Direction of Rural Electrification |
| PV Technology | Crystalline silicon |
| PV system supplier | Isofoton Maroc |
| Nominal power installed | 46 kWp |
| Type of PV system | Grid-connected |
| Operating date | June 14, 2007 |
| Duration of achievement | 4 months |
| Estimated PV system losses | 10 % |
| PV power production expected | 70 MWh per year which is the equivalent to the annual consumption of a village of 120 households (around 700 peoples) |
| PV power exploitation | Power production is exploited for domestic use in urban areas and the surplus is injected into the Moroccan network |
| Cost and financing | US $ 350,000 of which 140,000 funded by the German bank KFW |
| Estimated reduction of $CO_2$ emission | The micro-power should prevent the emission of 24 tons of carbon dioxide, or 720 tons over the project life. |

Table 1: Technical characteristics of the first Moroccan PV micro-power plant

## V- Results:

We have used the approach outlined above to estimate the annual PV performance of Chourouk micro-power plant. After providing the PVGIS software with all input parameters needed for performing the calculation, we obtained a large number of output results that we will expose and comment in the present section.

Even we have had the possibility to choose the inclination and orientation angle of Chourouk PV modules; we have preferred to let PVGIS calculating the optimal values for these parameters assuming fixed angles for the entire year. Table 2 summarizes the first results provided by PVGIS. We can see that the value of optimal inclination angle of the PV modules from the horizontal plane agrees perfectly with that of geographical latitude of Chourouk site. This is a good sign reflecting once more the accuracy of PVGIS calculation.

| Parameter | Value |
|---|---|
| Optimal inclination angle of the PV modules from the horizontal plane | 30° |
| Optimal orientation angle of the PV modules relative to the south direction | 0° (South) |
| Estimated losses due to temperature | 9.5 % (using local ambient temperature data) |
| Estimated losses due to angular reflectance effects | 2.6 % |
| Other losses (cables, inverter, etc.) | 10 % |
| Combined PV system losses | 22.1 % |

Table 2: Optimal inclination/orientation of Chourouk PV modules and their estimated losses

**V-1- Estimation of solar irradiation and daily temperature:**

Table 3 gives the estimated value of monthly ($G_m$) and daily ($G_d$) average sum of global irradiation per square meter received by the plane of Chourouk PV modules. It gives also the diffuse to global irradiation ratio D/G and the daily temperature $T_{24h}$ of the modules. The total yearly average value of the global irradiation at optimal inclination angle $G_{opt}$ has been found to be about 1980 kWh/m². The shadowing by local terrain features can affect theses values. Some of these results are displayed in figure 2. We can easily see that $G_m$ value of May is slightly higher than that of June. This is a little bit strange because usually the inverse should happen.

| Month | Daily average sum of global irradiation $G_d$ (kWh/m²) | Monthly average sum of global irradiation $G_m$ (kWh/m²) | Average ratio of Diffuse to Global irradiation D/G | Average value of daily Temperature $T_{24h}$ (°C) |
|---|---|---|---|---|
| Jan | 4.12 | 128 | 0.41 | 12.7 |
| Feb | 4.57 | 128 | 0.41 | 13.4 |
| Mar | 5.70 | 177 | 0.37 | 15.7 |
| Apr | 5.97 | 179 | 0.38 | 16.4 |
| May | 6.31 | 195 | 0.36 | 18.9 |
| Jun | 6.37 | 191 | 0.35 | 21.8 |
| Jul | 6.44 | 199 | 0.35 | 23.5 |
| Aug | 6.39 | 198 | 0.35 | 23.9 |
| Sep | 6.00 | 180 | 0.35 | 22.3 |
| Oct | 5.10 | 158 | 0.39 | 20.0 |
| Nov | 4.07 | 122 | 0.43 | 16.2 |
| Dec | 3.92 | 121 | 0.42 | 14.0 |
| **Average values** | **5.42** | **165** | **0.37** | **18.2** |

Table 3: PVGIS estimates of daily, monthly and yearly averages of the global irradiation at optimal inclination angle, the diffuse to global irradiation ratio and the daily temperature.

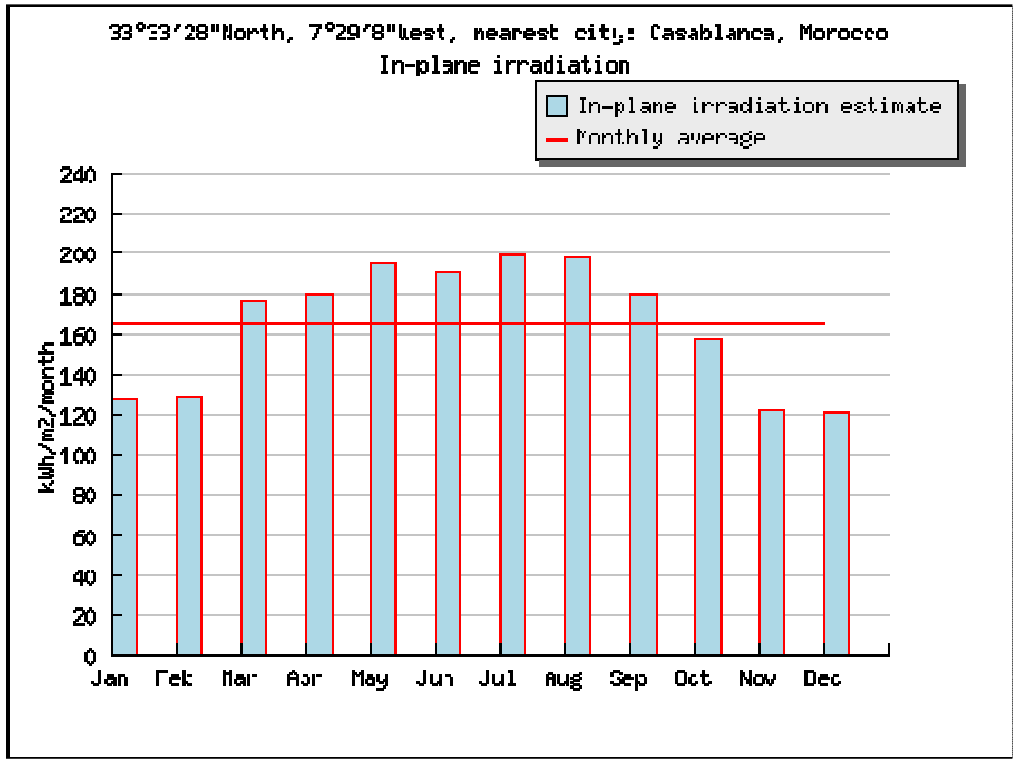

Figure 1: In-plane global irradiation estimated for the optimal inclination and orientation of Chourouk PV modules. The estimated monthly average value is also shown.

The values of irradiation and temperature obtained above are used to estimate and predict the amount of electricity production of Chourouk micro-power plant using the model and formulation described in the section II.

**V-2- Estimation of PV electricity production by Chourouk micro-power plant:**

Table 4 shows the estimated amount of electric power in kWh we can expect each day ($E_d$) and each month ($E_m$) from Chourouk micro-power plant with the parameters we inputted (at optimal inclination and orientation). It also shows the expected total PV power production par year. The electricity production for every month has been also plotted in figure 2 to show graphically the results. The power production ranges from about 4540 kWh in December to more than 7100 kWh in July. In December, the production is clearly much lower than other winter months because of the exceptionally rainy weather. The annual total estimated value of power production is about 71.875 MWh which is approximately the value initially expected and actually measured (70 MWh) as specified in table 1. The two values agree to better than 2.5%. This is within the typical 2 % uncertainty of the irradiance sensors and other error sources. The power production estimate for May is relatively higher than that for June. This unexpected result comes from the difference in $G_m$ values noticed above between these two months. We can therefore conclude that the PVGIS prediction is valid over a long period of time such as one year but it seems less appropriate for single months because of some discrepancies between estimated and expected values.

| Month | Average daily electricity production $E_d$ (kWh) | Average monthly electricity production $E_m$ (kWh) |
|---|---|---|
| Jan | 155 | 4799 |
| Feb | 171 | 4780 |
| Mar | 208 | 6452 |
| Apr | 218 | 6536 |
| May | 228 | 7059 |
| Jun | 228 | 6834 |
| Jul | 229 | 7108 |
| Aug | 228 | 7061 |
| Sep | 215 | 6452 |
| Oct | 185 | 5738 |
| Nov | 151 | 4516 |
| Dec | 146 | 4540 |
| **Average values** | **197** | **5990** |
| **Total annual production (kWh)** | colspan 71875 | |

Table 4: Estimated power production of Chourouk micro-power plant at optimal inclination and orientation

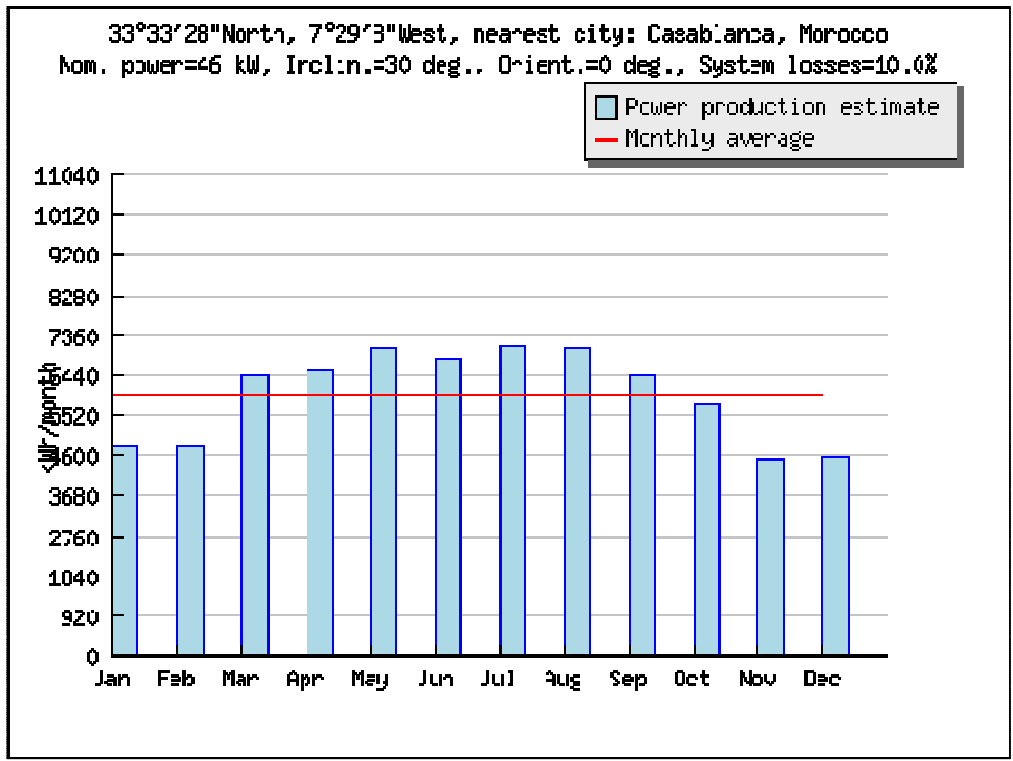

Figure 2: Estimated amount of electric power in kWh expected each month from Chourouk micr-power plant. The expected monthly average power production is also shown.

## VI- Conclusion:

In this paper, we applied PVGIS approach to the first Moroccan grid-connected micro-power PV plant recently built in Morocco with the aim to provide an analysis of in-site solar energy resources and PV electricity production from this station. For the optimal inclination and orientation angles determined by the software itself, we found that annual total power generation of the micro-power is only slightly higher than that initially expected at the installation stage and actually measured. The yearly predicted and measured power production values agree to about 2% which is the typical uncertainty of the irradiance sensors and other error sources. For two consecutive months which are May and June, unexpected results have been found both in term of solar energy resources and electricity production. We can therefore conclude that the PVGIS prediction is valid over a long period of time such as one year but it seems less appropriate for single months because of some discrepancies between estimated and expected values.

**Acknowledgements**

This paper has been finalized and written during the scientific stay of the first author A. Barhdadi, as Senior Associate Professor, at the Abdus Salam International Centre for Theoretical Physics (ICTP). This author would like to thank the Director and staff of the Centre for their efficient assistance and great support. He also wishes to thank Professor J. Niemela, head of ICTP Multidisciplinary Laboratory, for his scientific cooperation.